\documentstyle[12pt]{article}
\begin{document}\bibliographystyle{plain}
\begin{titlepage}\renewcommand{\thefootnote}{\fnsymbol{footnote}}
\hfill\begin{tabular}{l}HEPHY-PUB 714/99\\UWThPh-1999-30\\hep-ph/9905556\\May
1999\end{tabular}\\[2cm]
\Large\begin{center}{\bf THE ONE-DIMENSIONAL SPINLESS\\RELATIVISTIC COULOMB
PROBLEM}\\\vspace{2cm}\large{\bf Wolfgang LUCHA\footnote[1]{\normalsize\ {\em
E-mail\/}: wolfgang.lucha@oeaw.ac.at}}\\[.5cm]Institut f\"ur
Hochenergiephysik,\\\"Osterreichische Akademie der
Wissenschaften,\\Nikolsdorfergasse 18, A-1050 Wien, Austria\\[2cm]{\bf Franz
F. SCH\"OBERL\footnote[2]{\normalsize\ {\em E-mail\/}:
franz.schoeberl@univie.ac.at}}\\[.5cm]Institut f\"ur Theoretische
Physik,\\Universit\"at Wien,\\Boltzmanngasse 5, A-1090 Wien,
Austria\vfill{\normalsize\bf Abstract}\end{center}\small Motivated by a
recent analysis which presents explicitly the general solution, we
consider~the eigenvalue problem of the spinless Salpeter equation with a
(``hard-core amended'') Coulomb interaction potential in one dimension. We
prove the existence of a critical coupling constant (which contradicts the
assertions of the previous analysis) and give analytic upper bounds~on the
energy eigenvalues. These upper bounds seem to disprove the previous explicit
solution.

\vspace{3ex}

\noindent{\em PACS\/}: 03.65.Ge, 03.65.Pm
\renewcommand{\thefootnote}{\arabic{footnote}}\end{titlepage}

\normalsize

\section{Introduction: The Spinless Salpeter Equation in One Dimension}The
spinless Salpeter equation arises either as a standard reduction of the
well-known Bethe--Salpeter formalism \cite{SSE} for the description of bound
states within the framework of relativistic quantum field theory or as a
straightforward relativistic generalization~of the nonrelativistic
Schr\"odinger equation. This semirelativistic equation of motion with a
static interaction described by the Coulomb potential (originating, for
instance, from the exchange of a massless particle between the bound-state
constituents) defines what we call, for short, the ``spinless relativistic
Coulomb problem.\footnote{\normalsize\ The present state-of-the-art of the
three-dimensional relativistic Coulomb problem has been reviewed, for
instance, in Refs.~\cite{Lucha94:C,Lucha98:R,Lucha98:T}.}''

Recently, confining the configuration space to the positive half-line (and
mimicking thereby the effect of a ``hard-core''), the relativistic Coulomb
problem has been studied in one dimension \cite{Brau99}. This one-dimensional
case may serve as a toy model which might prove to be instructive for the
analysis of the still unsolved three-dimensional problem. In view of its
potential importance, we re-analyze this nontrivial and delicate problem.

The spinless Salpeter equation may be regarded as the eigenvalue equation
$$H|\chi_k\rangle=E_k|\chi_k\rangle\ ,\quad k=1,2,3,\dots\ ,$$for the
complete set of Hilbert-space eigenvectors $|\chi_k\rangle$ and corresponding
eigenvalues
$$E_k\equiv\frac{\langle\chi_k|H|\chi_k\rangle}{\langle\chi_k|\chi_k\rangle}$$
of a self-adjoint operator $H$ of Hamiltonian form, consisting of a
momentum-dependent kinetic-energy operator and a coordinate-dependent
interaction-potential operator:\begin{equation}H=T+V\
,\label{Eq:SRH}\end{equation}where $T$ is the ``square-root'' operator of the
relativistic kinetic energy of some particle of mass $m$ and momentum
$p$,\begin{equation}T=T(p)\equiv\sqrt{p^2+m^2}\
,\label{eq:RKE}\end{equation}and $V=V(x)$ is an arbitrary,
coordinate-dependent, static interaction potential. The action of the
kinetic-energy operator $T$ on an element $\psi$ of $L_2(R)$, the Hilbert
space~of square-integrable functions on the real line, $R$, is defined by
(cf. also Eq.~(3) of Ref.~\cite{Brau99})
\begin{equation}(T\psi)(x)=\frac{1}{2\pi}\int\limits_{-\infty}^{+\infty} {\rm
d}p\int\limits_{-\infty}^{+\infty}{\rm d}y\,\sqrt{p^2+m^2}\,\exp[{\rm
i}\,p\,(x-y)]\,\psi(y)\ .\label{Eq:AOT}\end{equation}

In Ref.~\cite{Brau99}, the domain of $H$ is restricted to square-integrable
functions $\Psi(x)$ with support on the positive real line, $R^+$, only,
vanishing at $x=0$ (cf. Eq.~(28) of Ref.~\cite{Brau99}):
$$\Psi(x)=0\quad\mbox{for}\ x\le0\ .$$This restriction may be interpreted as
due to the presence of a ``hard-core'' interaction potential effective for
$x\le0$. For $x>0$, the interaction potential $V$ is chosen to be of Coulomb
type, its strength parametrized by a positive coupling constant $\alpha$,
i.e.,~$\alpha>0$:$$V(x)=V_{\rm C}(x)=-\frac{\alpha}{x}\quad\mbox{for}\ x>0\
.$$Let the Coulomb-type semirelativistic Hamiltonian $H_{\rm C}$ be the
operator defined in this way.

\section{Concerns --- Dark Clouds Appear at the Horizon}Now, according to the
analysis of Ref.~\cite{Brau99}, the point spectrum of the Hamiltonian $H_{\rm
C}$ consists of the set of eigenvalues (cf. Eq.~(33) of
Ref.~\cite{Brau99})\begin{equation}\widetilde
E_n=\frac{m}{\displaystyle\sqrt{1+\frac{\alpha^2}{n^2}}}\ ,\quad
n=1,2,3,\dots\ .\label{Eq:EEVs}\end{equation}The corresponding eigenfunctions
$\Psi_n(x)$ must be of the form (cf. Eq.~(28) of
Ref.~\cite{Brau99})\begin{equation}\Psi_n(x)=\psi_n(x)\,\Theta(x)\ ,\quad
n=1,2,3,\dots\ ,\label{Eq:HEF}\end{equation}where $\Theta(x)$ denotes the
Heaviside step function, defined here
by\begin{eqnarray*}\Theta(x)&=&1\quad\mbox{for}\ x>0\
,\\[1ex]\Theta(x)&=&0\quad\mbox{for}\ x\le0\ .\end{eqnarray*}In particular,
the (not normalized) eigenfunctions $\psi_n(x)$, $n=1,2,3,$ corresponding to
the lowest energy eigenvalues $\widetilde E_n$ are explicitly given by (cf.
Eqs.~(37)--(40) of Ref.~\cite{Brau99})\begin{eqnarray}
\psi_1(x)&=&x\,\exp(-\beta_1\,x)\ ,\nonumber\\[1ex]
\psi_2(x)&=&x\left(x-\frac{m^2}{S_2^2\,\beta_2}\right)\exp(-\beta_2\,x)\
,\nonumber\\[1ex]
\psi_3(x)&=&x\left(x^2-\frac{3\,m^2}{S_3^2\,\beta_3}\,x+
\frac{3\,m^2\,(\beta_3^2+m^2)}{2\,S_3^4\,\beta_3^2}\right)\exp(-\beta_3\,x)\
,\label{Eq:LEF}\end{eqnarray}with (cf. Eq.~(32) of
Ref.~\cite{Brau99})$$\beta_n\equiv
\frac{m\,\alpha}{n\,\displaystyle\sqrt{1+\frac{\alpha^2}{n^2}}}=
\frac{\alpha}{n}\,\widetilde E_n\ ,\quad n=1,2,3,\dots\ ,$$and the
abbreviation (cf. Eq.~(26) of
Ref.~\cite{Brau99})$$S_n\equiv\sqrt{m^2-\beta_n^2}\ ,\quad n=1,2,3,\dots\ .$$

However, there are some facts which cause severe doubts about the validity of
this solution:\begin{description}\item[Boundedness from below:]For coupling
constants $\alpha$ larger than some critical value $\alpha_{\rm c}$ (which
has yet to be determined), the operator $H_{\rm C}$ is not bounded from
below. This may be seen, for instance, already from the expectation value of
$H_{\rm C}$ with respect to the (normalized) trial state $|\Phi\rangle$
defined by the configuration-space~trial function
$$\Phi(x)=\varphi(x)\,\Theta(x)$$with
$$\varphi(x)=2\,\mu^{3/2}\,x\,\exp(-\mu\,x)\ ,\quad\mu>0\ ,$$and satisfying
the normalization condition$$\||\Phi\rangle\|^2\equiv\langle\Phi|\Phi\rangle=
\int\limits_0^\infty{\rm d}x\,|\varphi(x)|^2=1\ .$$Apart from the
arbitrariness of the variational parameter $\mu$, this trial function~$\Phi$
coincides, in fact, with the ground-state solution $\Psi_1$ as given in
Eqs.~(\ref{Eq:HEF}), (\ref{Eq:LEF}).~The expectation value of the Coulomb
interaction-potential operator $V_{\rm C}$ with respect to the trial state
$|\Phi\rangle$ reads$$\langle\Phi|V_{\rm
C}|\Phi\rangle=-\alpha\int\limits_0^\infty{\rm
d}x\,\frac{1}{x}|\varphi(x)|^2=-\mu\,\alpha\ .$$There is a trivial (but
nevertheless fundamental) inequality for the expectation values of a
self-adjoint (but otherwise arbitrary) operator ${\cal O}={\cal O}^\dagger$
and its square, taken with respect to an arbitrary Hilbert-space state
$|\psi\rangle$ in the domain ${\cal D}({\cal O})$~of this operator ${\cal
O}$:$$\frac{|\langle\psi|{\cal
O}|\psi\rangle|}{\langle\psi|\psi\rangle}\le\sqrt{\frac{\langle\psi|{\cal
O}^2|\psi\rangle}{\langle\psi|\psi\rangle}}\quad\mbox{for all}\
|\psi\rangle\in{\cal D}({\cal O})\ .$$Application of this inequality to the
kinetic-energy operator $T$ of Eq.~(\ref{eq:RKE}) allows to get rid of the
troublesome square-root operator:$$\langle\Phi|T|\Phi\rangle\le
\sqrt{\langle\Phi|T^2|\Phi\rangle}\equiv\sqrt{\langle\Phi|p^2|\Phi\rangle+m^2}\
.$$The expectation value of $p^2$ required here
reads$$\langle\Phi|p^2|\Phi\rangle=\mu^2\ .$$Thus the expectation value of
the Coulomb-like semirelativistic Hamiltonian~$H_{\rm C}$ with respect to the
trial state $|\Phi\rangle$ is bounded from above by
\begin{equation}\langle\Phi|H_{\rm C}|\Phi\rangle=
\langle\Phi|T+V_{\rm C}|\Phi\rangle\le\sqrt{\mu^2+m^2}-\mu\,\alpha\
.\label{Eq:EVOCH}\end{equation}When inspecting this inequality in the limit
of large $\mu$, that is, for $\mu\to\infty$, one realizes that, for $\alpha$
large enough, the operator $H_{\rm C}$ is not bounded from below.~In fact,
the expectation value of the kinetic-energy operator $T$ with respect to the
trial state $|\Phi\rangle$,\begin{equation}\langle\Phi|T|\Phi\rangle
=\int\limits_{-\infty}^{+\infty}{\rm d}x\,\Phi^\ast(x)\,(T\Phi)(x)
=\frac{4\,\mu^3}{\pi}\int\limits_0^\infty{\rm d}p\,
\frac{\displaystyle\sqrt{p^2+m^2}}{(p^2+\mu^2)^2}\
,\label{Eq:EVKE}\end{equation}is simple enough to be investigated explicitly.
For $\mu\gg m$, this expectation value simplifies
to$$\langle\Phi|T|\Phi\rangle=\frac{2\,\mu}{\pi}\quad\mbox{for}\ \mu\gg m\
.$$Consequently, in the (ultrarelativistic) limit $\mu\to\infty$, the
expectation value of~$H_{\rm C}$ behaves
like$$\lim_{\mu\to\infty}\frac{\langle\Phi|H_{\rm
C}|\Phi\rangle}{\mu}=\frac{2}{\pi}-\alpha\ .$$This clearly indicates that for
the Hamiltonian $H_{\rm C}$ to be bounded from below~the Coulomb coupling
constant $\alpha$ has to be bounded from above by the critical
value$$\alpha_{\rm c}\le\frac{2}{\pi}\ .$$(This upper bound on $\alpha_{\rm
c}$ is, in fact, identical to the critical coupling constant $\alpha_{\rm c}$
found in the case of the three-dimensional spinless relativistic Coulomb
problem \cite{Herbst}.)\item[Upper bound on lowest eigenvalue:]As rather
trivial consequence of the famous minimum--maximum principle \cite{MMP}, the
expectation
value$$\frac{\langle\psi|H|\psi\rangle}{\langle\psi|\psi\rangle}$$of a
self-adjoint operator $H$ bounded from below, with respect to some arbitrary
state $|\psi\rangle$ in the domain of $H$, ${\cal D}(H)$, is always larger
than or equal to the lowest eigenvalue $E_1$ of $H$:\footnote{\normalsize\
This statement constitutes what is sometimes simply called ``Rayleigh's
principle.''}$$E_1\le
\frac{\langle\psi|H|\psi\rangle}{\langle\psi|\psi\rangle}\quad\mbox{for all}\
|\psi\rangle\in{\cal D}(H)\ .$$Accordingly, minimizing the expression on the
right-hand side of inequality (\ref{Eq:EVOCH}) with respect to the
variational parameter $\mu$ yields a simple analytic upper bound $\widehat
E_1$ on the ground-state energy eigenvalue $E_1$ of the Coulomb-like
semirelativistic Hamiltonian $H_{\rm C}$:$$E_1\le\widehat
E_1$$with\begin{equation}\widehat E_1=m\,\sqrt{1-\alpha^2}\
.\label{Eq:AUBGSE}\end{equation}The same analytic upper bound on the
ground-state energy $E_1$ has been found~in the case of the three-dimensional
spinless relativistic Coulomb problem
\cite{Lucha94varbound,Lucha96:AUB,Lucha98:T}. Reality of this latter
expression requires again the existence of a critical coupling constant
$\alpha_{\rm c}$ and indicates that this critical value of $\alpha$ is less
than or equal to 1:$$\alpha_{\rm c}\le1\ .$$Moreover, at least for the energy
eigenvalue $E_1$ corresponding to the ground~state of the Hamiltonian $H_{\rm
C}$, the supposedly {\em exact\/} value of Eq.~(\ref{Eq:EEVs}),
\begin{equation}\widetilde E_1=\frac{m}{\displaystyle\sqrt{1+\alpha^2}}\
,\label{Eq:GSE}\end{equation}is in clear conflict with the naive {\em upper
bound\/} $\widehat E_1$ of Eq.~(\ref{Eq:AUBGSE}):$$\frac{\widehat
E_1}{\widetilde E_1}=\sqrt{1-\alpha^4}$$and therefore$$\widehat
E_1<\widetilde E_1\quad\mbox{for}\ \alpha>0\ .$$For larger values of the
Coulomb coupling constant $\alpha$, the upper bound (\ref{Eq:AUBGSE}) on the
ground-state energy can be easily improved by fixing in the
expectation~value~(\ref{Eq:EVKE}) of the kinetic-energy operator $T$ the
variational parameter $\mu$ to the value $\mu=m$. In this case, this
expectation value reads$$\langle\Phi|T|\Phi\rangle=\frac{4\,m}{\pi}\
.$$Accordingly, the ground-state energy eigenvalue $E_1$ is bounded from
above by\begin{equation}E_1\le\left(\frac{4}{\pi}-\alpha\right)m\
.\label{Eq:AUBGSE-mu=m}\end{equation}For the Coulomb coupling constant
$\alpha$ in the range
$$\frac{2}{\pi}-\sqrt{\frac{1}{2}-\frac{4}{\pi^2}}<\alpha\le\frac{2}{\pi}\
,$$the above expression represents a genuine improvement of the upper bound
(\ref{Eq:AUBGSE}).\item[Eigenstate expectation values vs. eigenvalues:] The
expectation value (\ref{Eq:EVKE}) of~the kinetic-energy operator $T$ with
respect to the trial state $|\Phi\rangle$ may be written~down
explicitly:$$\langle\Phi|T|\Phi\rangle=\frac{2}{\pi}\,m\left[\frac{\mu}{m}
+\displaystyle\frac{\arccos\displaystyle\frac{\mu}{m}}
{\displaystyle\sqrt{1-\frac{\mu^2}{m^2}}}\right].$$Now,
for$$\mu=\beta_1=\frac{m\,\alpha}{\displaystyle\sqrt{1+\alpha^2}}\ ,$$the
trial function $\Phi$ coincides with the normalized ground-state
eigenfunction~$\Psi_1$. In this case, the corresponding expectation value of
the Hamiltonian $H_{\rm C}$ becomes\begin{equation}\langle\Psi_1|H_{\rm
C}|\Psi_1\rangle=
\frac{m}{\displaystyle\sqrt{1+\alpha^2}}\left[\frac{2}{\pi}\left(\alpha+
(1+\alpha^2)\,\mbox{arccot}\,\alpha\right)-\alpha^2\right].
\label{Eq:GSE-expl}\end{equation}Unfortunately, the above expectation value
does not agree with the ground-state energy (\ref{Eq:GSE}) deduced from
Eq.~(\ref{Eq:EEVs}):$$\langle\Psi_1|H_{\rm C}|\Psi_1\rangle\ne\widetilde E_1\
.$$\item[Orthogonality of eigenstates:]Eigenstates $|\chi_i\rangle$,
$i=1,2,3,\dots,$ of some self-adjoint operator $H$ corresponding to distinct
eigenvalues of $H$ are mutually
orthogonal:$$\langle\chi_i|\chi_k\rangle\propto\delta_{ik}\ ,\quad
i,k=1,2,3,\dots\ .$$This feature is definitely not exhibited by the
overlaps$$\langle\Psi_i|\Psi_k\rangle=\int\limits_{-\infty}^{+\infty}{\rm
d}x\,\Psi_i^\ast(x)\,\Psi_k(x)=\int\limits_0^\infty{\rm
d}x\,\psi_i^\ast(x)\,\psi_k(x)\ ,\quad i,k=1,2,3,\dots\ ,$$of the lowest
eigenfunctions $\Psi_i(x)$, $i=1,2,3,$ given in Eqs.~(\ref{Eq:HEF}),
(\ref{Eq:LEF}). For instance, the overlap $\langle\Psi_1|\Psi_2\rangle$ of
the ground state $|\Psi_1\rangle$ and the first excitation $|\Psi_2\rangle$
is given by$$\langle\Psi_1|\Psi_2\rangle=
\frac{2\,[3\,S_2^2\,\beta_2-m^2\,(\beta_1+\beta_2)]}
{(\beta_1+\beta_2)^4\,S_2^2\,\beta_2}\ ,$$revealing thus, beyond doubt, the
non-orthogonality of the vectors $|\Psi_1\rangle$
and~$|\Psi_2\rangle$.\end{description}

\section{Exact Analytic Upper Bounds on Energy Levels}In view of the above,
let us try to collect unambiguous results for the one-dimensional spinless
relativistic Coulomb problem. With the help of the definition (\ref{Eq:AOT})
of the action of a momentum-dependent operator in coordinate space, it is
easy to convince oneself of the validity of the operator inequality$$T\le
T_{\rm NR}\equiv m+\frac{p^2}{2\,m}\ ;$$the relativistic kinetic-energy
operator $T$ is bounded from above by its nonrelativistic counterpart $T_{\rm
NR}$: when introducing the Fourier transform $\widetilde\psi(p)$ of the
coordinate-space representation $\psi(x)$ of the Hilbert-space vector
$|\psi\rangle$,$$\widetilde\psi(p)\equiv\frac{1}{\sqrt{2\pi}}
\int\limits_{-\infty}^{+\infty} {\rm d}x\,\exp[-{\rm i}\,p\,x]\,\psi(x)\
,$$one finds\begin{eqnarray*}\langle\psi|T_{\rm NR}-T|\psi\rangle&=&
\int\limits_{-\infty}^{+\infty}{\rm d}x\,\psi^\ast(x)\left[(T_{\rm
NR}\psi)(x)-(T\psi)(x)\right]\\[1ex]&=&
\int\limits_{-\infty}^{+\infty}{\rm d}p\,|\widetilde\psi(p)|^2\left(
m+\frac{p^2}{2\,m}-\sqrt{p^2+m^2}\right)\\[1ex]&\ge&0\ .\end{eqnarray*}
Hence, adding the Coulomb interaction potential $V_{\rm C}$, the
semirelativistic Hamiltonian $H_{\rm C}$ is, of course, bounded from above by
the corresponding nonrelativistic Hamiltonian $H_{\rm C,NR}$:$$H_{\rm C}\le
H_{\rm C,NR}\equiv T_{\rm NR}+V_{\rm C}\ .$$Now, upon invoking the
minimum--maximum principle \cite{MMP} (which requires the operator $H_{\rm
C}$ to be both self-adjoint and bounded from below) and combining this
principle with the above operator inequality, we infer that every eigenvalue
$E_n$, $n=1,2,3,\dots,$ of~$H_{\rm C}$ is bounded from above by a
corresponding eigenvalue $E_{n,\rm NR}$, $n=1,2,3,\dots,$ of $H_{\rm
C,NR}$:\footnote{\normalsize\ The line of arguments leading to the general
form of this statement may be found, for instance, in
Refs.~\cite{Lucha96:AUB,Lucha98:T}. It is summarized in
Appendix~\ref{App:MMPOI}. For a rather brief account of the application of
these ideas to the three-dimensional spinless relativistic Coulomb problem,
see, e.g., Ref.~\cite{Lucha98:R}.}$$E_n\le E_{n,\rm NR}\quad\mbox{for}\
n=1,2,3,\dots\ .$$It is a simple and straightforward exercise to calculate
the latter set of eigenvalues:$$E_{n,\rm
NR}=m\left(1-\frac{\alpha^2}{2\,n^2}\right),\quad n=1,2,3,\dots\ .$$These
upper bounds on the energy eigenvalues $E_n$ may be easily improved by the
same reasoning as before. Introducing an arbitrary real parameter $\eta$
(with the dimension~of mass), we find a set of operator inequalities for the
kinetic energy $T$ \cite{Lucha96:AUB,Lucha98:T,Martin},~namely,
$$T\le\frac{p^2+m^2+\eta^2}{2\,\eta}\quad\mbox{for all}\ \eta>0$$and,
consequently, a set of operator inequalities for the Coulomb-type
semirelativistic Hamiltonian $H_{\rm C}$ \cite{Lucha96:AUB,Lucha98:T}:
$$H_{\rm C}\le\widehat H_{\rm
C}(\eta)\equiv\frac{p^2+m^2+\eta^2}{2\,\eta}+V_{\rm C}\quad\mbox{for all}\
\eta>0\ .$$Accordingly, every eigenvalue $E_n$, $n=1,2,3,\dots,$ of $H_{\rm
C}$ is bounded from above by~the minimum, with respect to the mass parameter
$\eta$, of the corresponding eigenvalue \cite{Lucha96:AUB,Lucha98:T}
$$\widehat E_{n,\rm C}(\eta)=\frac{1}{2\,\eta}\left[m^2+\eta^2
\left(1-\frac{\alpha^2}{n^2}\right)\right],\quad n=1,2,3,\dots\ ,$$of
$\widehat H_{\rm C}(\eta)$:$$E_n\le\displaystyle\min_{\eta>0}\widehat
E_{n,\rm C}(\eta)=m\,\sqrt{1-\frac{\alpha^2}{n^2}}\quad\mbox{for all}\
\alpha\le\alpha_{\rm c}\ .$$For $n=1$, this (variational) upper bound
coincides with the previous upper bound~(\ref{Eq:AUBGSE}). It goes without
saying that these upper bounds are violated by the energy eigenvalues
$\widetilde E_n$ given in Eq.~(\ref{Eq:EEVs}):$$\frac{1}{\widetilde
E_n}\,\displaystyle\min_{\eta>0}\widehat E_{n,\rm
C}(\eta)=\sqrt{1-\frac{\alpha^4}{n^4}}<1\quad\mbox{for}\ \alpha\ne 0\
,\quad\mbox{for all}\ n=1,2,3,\dots\
,$$means$$\displaystyle\min_{\eta>0}\widehat E_{n,\rm C}(\eta)<\widetilde
E_n\quad\mbox{for}\ \alpha\ne 0\ ,\quad\mbox{for all}\ n=1,2,3,\dots\ !$$

Moreover, for $\mu=m\,\alpha$, our generic trial state $|\Phi\rangle$ becomes
the lowest eigenstate~of the nonrelativistic Hamiltonian $H_{\rm C,NR}$,
corresponding to the ground-state eigenvalue\footnote{\normalsize\ The
Coulomb problem involves no dimensional parameter other than the particle
mass $m$. Therefore, both the energy eigenvalues $E_n$ and the parameter(s)
$\mu$ have to~be proportional to $m$.}$$E_{1,\rm
NR}=m\left(1-\frac{\alpha^2}{2}\right),$$which may be easily seen:$$(T_{\rm
NR}\varphi)(x)=\left(m-\frac{1}{2\,m}\,\frac{{\rm d}^2}{{\rm
d}x^2}\right)\varphi(x)=
\left(m-\frac{\mu^2}{2\,m}+\frac{\mu}{m}\,\frac{1}{x}\right)\varphi(x)\quad
\mbox{for}\ x>0$$implies (with $\mu=m\,\alpha$)$$H_{\rm
C,NR}|\Phi\rangle=E_{1,\rm NR}|\Phi\rangle\ .$$It appears rather unlikely
that the same functional form represents also the eigenstate of the
semirelativistic Hamiltonian $H_{\rm C}$.

\section{Summary, Further Considerations, Conclusions}This work is devoted to
the study of the one-dimensional spinless relativistic Coulomb problem on the
positive half-line. Assuming a (dense) domain in $L_2(R^+)$ such that~the
semirelativistic Coulombic Hamiltonian $H_{\rm C}$ defined in the
Introduction is self-adjoint, analytic upper bounds on the energy eigenvalues
$E_k$, $k=1,2,3,\dots,$ have been derived:\begin{equation}E_k\le
m\,\sqrt{1-\frac{\alpha^2}{k^2}}\quad\mbox{for all}\ k=1,2,3,\dots\
.\label{Eq:AUBOELFSRCP}\end{equation}Surprisingly, the explicit solution
presented in Ref.~\cite{Brau99} does not fit into these bounds.

\newpage In order to cast some light into this confusing situation, let us
inspect the action~(\ref{Eq:AOT}) of the kinetic-energy operator $T$ in more
detail. Consider not normalized Hilbert-space vectors $|\Phi_n\rangle$,
$n=0,1,2,\dots,$ defined, as usual, by the coordinate-space representation
$$\Phi_n(x)=x^n\,\exp(-\mu\,x)\,\Theta(x)\ ,\quad\mu>0\ ,\quad n=0,1,2,\dots\
.$$These vectors certainly belong to the Hilbert space $L_2(R)$ for all
$n=0,1,2,\dots,$
since$$\||\Phi_n\rangle\|^2\equiv\langle\Phi_n|\Phi_n\rangle=
\int\limits_{-\infty}^{+\infty}{\rm d}x\,|\Phi_n(x)|^2=
\int\limits_0^\infty{\rm d}x\,x^{2\,n}\,\exp(-2\,\mu\,x)
=\frac{\Gamma(2\,n+1)}{(2\,\mu)^{2\,n+1}}<\infty\ .$${\em However\/}: The
norm $\|T|\Phi_n\rangle\|$ of the vectors $T|\Phi_n\rangle$, $n=0,1,2,\dots,$
may be found~from$$\|T|\Phi_n\rangle\|^2=\int\limits_{-\infty}^{+\infty}{\rm
d}x\,|(T\Phi_n)(x)|^2=
\frac{[\Gamma(n+1)]^2}{2\pi}\int\limits_{-\infty}^{+\infty}{\rm
d}p\,\frac{p^2+m^2}{(p^2+\mu^2)^{n+1}}\ .$$This observation might be a hint
that the vector $|\Phi_0\rangle$, that is,
$\Phi_0(x)=\exp(-\mu\,x)\,\Theta(x)$, does {\em not\/} belong to the domain
of the kinetic-energy operator $T$. If this is indeed true, it is by no means
obvious how to make sense of Eq.~(16) of Ref.~\cite{Brau99} for the case
$n=0$.

Trivially, {\em if\/} Eq.~(16) of Ref.~\cite{Brau99} is correct for $n=0$,
all these relations for arbitrary $n=1,2,\dots$ may be obtained by a simple
differentiation of the relation for $n=0$~with respect to the (generic)
parameter $\mu$, taking advantage of$$T\,x^n\,\exp(-\mu\,x)=\left(-\frac{{\rm
d}}{{\rm d}\mu}\right)^n\,T\,\exp(-\mu\,x)\ .$$

Similarly, it is somewhat hard to believe that Eq.~(16) of Ref.~\cite{Brau99}
holds for $n=1$.~In our notation, Eq.~(16) of Ref.~\cite{Brau99} {\em
would\/} read for
$n=1$$$(T\Phi_1)(x)=\left(S+\frac{\mu}{S\,x}\right)\Phi_1(x)$$with
$$S\equiv\sqrt{m^2-\mu^2}\ .$$Considering merely the norms of the vectors on
both sides of this equation, we find,~for the norm of the vector on the
left-hand side,$$\|T|\Phi_1\rangle\|^2=\frac{m^2+\mu^2}{4\,\mu^3}$$but, for
the norm of the vector on the right-hand
side,$$\left\|\left(S+\frac{\mu}{S\,x}\right)|\Phi_1\rangle\right\|^2=
\frac{m^4+\mu^4}{4\,\mu^3\,S^2}\ .$$These two expressions for the norms
become equal only for the---excluded---case $\mu=0$. Unfortunately, precisely
the above relation forms the basis for the assertion in Ref.~\cite{Brau99}
that $\Phi_1(x)$ with $\mu=\beta_1$ is the ground-state eigenfunction of the
(``hard-core amended'') one-dimensional spinless relativistic Coulomb problem
as defined in the Introduction.

In conclusion, let us summarize our point of view as follows: The energy
eigenvalues $E_k$, $k=1,2,3,\dots,$ of the one-dimensional spinless
relativistic Coulomb problem (with hard-core interaction on the nonpositive
real line) are bounded from above by~Eq.~(\ref{Eq:AUBOELFSRCP}). For the
ground-state energy eigenvalue $E_1$, this upper bound may be improved
to~some extent, by considering appropriately the minimum of the bounds of
Eq.~(\ref{Eq:AUBGSE-mu=m}), Eq.~(\ref{Eq:GSE-expl}), or
Eq.~(\ref{Eq:AUBOELFSRCP}) for $k=1$, that is, Eq.~(\ref{Eq:AUBGSE}). To our
knowledge, these upper bounds represent the only information available at
present about the exact location of the energy levels of the (``hard-core
amended'') one-dimensional spinless relativistic Coulomb problem.

\section*{Acknowledgements}We would like to thank H.~Narnhofer for
stimulating discussions and a critical reading of the manuscript.

\appendix\section[]{Combining Minimum--Maximum Principle with Operator
Inequalities \cite{Lucha96:AUB,Lucha98:T}}\label{App:MMPOI}There exist
several equivalent formulations of the well-known ``min--max
principle''~\cite{MMP}. For practical purposes, the most convenient one is
perhaps the following:\begin{itemize}\item Let $H$ be a self-adjoint operator
bounded from below.\item Let $E_k$, $k=1,2,3,\dots,$ denote the eigenvalues
of $H$, defined by$$H|\chi_k\rangle=E_k|\chi_k\rangle\ ,\quad k=1,2,3,\dots\
,$$and ordered according to$$E_1\le E_2\le E_3\le\dots\ .$$\item Consider
only the eigenvalues $E_k$ below the onset of the essential spectrum
of~$H$.\item Let $D_d$ be some $d$-dimensional subspace of the domain ${\cal
D}(H)$ of $H$: $D_d\subset{\cal D}(H)$.\end{itemize} Then the $k$th
eigenvalue $E_k$ (when counting multiplicity) of $H$ satisfies the inequality
$$E_k\le\displaystyle\sup_{|\psi\rangle\in D_k}
\frac{\langle\psi|H|\psi\rangle}{\langle\psi|\psi\rangle}\quad\mbox{for}\
k=1,2,3,\dots\ .$$The min--max principle may be employed in order to compare
eigenvalues of operators:\begin{itemize}\item Assume the validity of a
generic operator inequality of the form$$H\le{\cal O}\ .$$Then$$E_k\equiv
\frac{\langle\chi_k|H|\chi_k\rangle}{\langle\chi_k|\chi_k\rangle}
\le\sup_{|\psi\rangle\in D_k}
\frac{\langle\psi|H|\psi\rangle}{\langle\psi|\psi\rangle}
\le\sup_{|\psi\rangle\in D_k}\frac{\langle\psi|{\cal
O}|\psi\rangle}{\langle\psi|\psi\rangle}\ .$$\item Assume that the
$k$-dimensional subspace $D_k$ in this inequality is spanned by the first $k$
eigenvectors of the operator ${\cal O}$, that is, by precisely those
eigenvectors of~${\cal O}$ that correspond to the first $k$ eigenvalues
$\widehat E_1,\widehat E_2,\dots,\widehat E_k$ of ${\cal O}$ if the
eigenvalues~of ${\cal O}$ are ordered according to$$\widehat E_1\le\widehat
E_2\le\widehat E_3\le\dots\ .$$Then$$\sup_{|\psi\rangle\in
D_k}\frac{\langle\psi|{\cal O}|\psi\rangle}{\langle\psi|\psi\rangle}=\widehat
E_k\ .$$\end{itemize}Consequently, every eigenvalue $E_k$, $k=1,2,3,\dots,$
of $H$ is bounded from above by~the corresponding eigenvalue $\widehat E_k$,
$k=1,2,3,\dots,$ of ${\cal O}$:$$E_k\le\widehat E_k\quad\mbox{for}\
k=1,2,3,\dots\ .$$


\begin{thebibliography}{30}
\bibitem{SSE}E.~E.~Salpeter and H.~A.~Bethe, Phys.~Rev. {\bf 84} (1951)
1232;\\E.~E.~Salpeter, Phys.~Rev. {\bf 87} (1952) 328.
\bibitem{Lucha94:C}W.~Lucha and F.~F.~Sch\"oberl, in: {\em Proceedings of the
International Conference on Quark Confinement and the Hadron Spectrum\/}
(Como, Italy, June 1994), eds. N. Brambilla and G.~M.~Prosperi (World
Scientific, River Edge, N.~J., 1995) p.~100, hep-ph/9410221.
\bibitem{Lucha98:R}W.~Lucha and F.~F.~Sch\"oberl, HEPHY-PUB 692/98 (1998),
hep-ph/9807342.
\bibitem{Lucha98:T}W.~Lucha and F.~F.~Sch\"oberl, HEPHY-PUB 701/98 (1998),
hep-ph/9812368, Int. J. Mod. Phys. A (in print);\\W.~Lucha and
F.~F.~Sch\"oberl, HEPHY-PUB 706/98 (1998), hep-ph/9812526, to appear in the
Proceedings of the International Conference on ``Nuclear \&~Particle Physics
with CEBAF at Jefferson Lab,'' Nov. 3 -- 10, 1998, Dubrovnik, Croatia.
\bibitem{Brau99}F.~Brau, J.~Math.~Phys.\ {\bf 40} (1999) 1119.
\bibitem{Herbst}I.~W.~Herbst, Commun. Math. Phys. {\bf 53} (1977) 285; {\bf
55} (1977) 316 (addendum).
\bibitem{MMP}M.~Reed and B.~Simon, {\em Methods of Modern Mathematical
Physics~IV: Analysis~of Operators\/} (Academic Press, New York, 1978)
Section~XIII.1;\\A.~Weinstein and W.~Stenger, {\em Methods of Intermediate
Problems for Eigenvalues -- Theory and Ramifications} (Academic Press, New
York/London, 1972) Chapters 1 and 2;\\W.~Thirring, {\em A Course in
Mathematical Physics 3: Quantum Mechanics of Atoms and Molecules\/}
(Springer, New York/Wien, 1990) Section~3.5.
\bibitem{Lucha94varbound}W.~Lucha and F.~F.~Sch\"oberl, Phys.~Rev.~D {\bf 50}
(1994) 5443, hep-ph/9406312.
\bibitem{Lucha96:AUB}W.~Lucha and F.~F.~Sch\"oberl, Phys.~Rev.~A {\bf 54}
(1996) 3790, hep-ph/9603429.
\bibitem{Martin}A.~Martin and S.~M.~Roy, Phys.~Lett.~B {\bf 233} (1989) 407.
\end{thebibliography}
\end{document}